\documentclass [aps,prb,preprint,preprintnumbers,amsmath,amssymb,superscriptaddress]{revtex4}

\usepackage{graphics}
\usepackage{graphicx}
\usepackage{epsfig}

\newcommand{\beq}{\begin{equation}}
\newcommand{\eeq}{\end{equation}}

\newcommand{\bea}{\begin{eqnarray}}
\newcommand{\eea}{\end{eqnarray}}
\usepackage{dcolumn}
\usepackage{bm}
\begin{document}

\title{Hyperfine Interactions and Spin Transport in Ferromagnet-Semiconductor Heterostructures}

\author{M. K. Chan}
\affiliation{School of Physics and Astronomy, University of
Minnesota, Minneapolis, MN 55455}
\author{Q. O. Hu}
\affiliation{Departments of
Electrical and Computer Engineering and Materials, University of
California, Santa Barbara, CA 93106}
\author{J. Zhang}
\affiliation{School of Physics and Astronomy, University of
Minnesota, Minneapolis, MN 55455}
\author{T. Kondo}
\altaffiliation{On leave from Tokyo Institute of Technology,
Yokohama, Kanagawa 226-8503, Japan.  Current address:  Corporate
R\&D center, Toshiba Corp., Kawasaki, Japan} \affiliation{School of
Physics and Astronomy, University of Minnesota, Minneapolis, MN
55455}
\author {C. J. Palmstr\o m}
\affiliation{Departments of
Electrical and Computer Engineering and Materials, University of
California, Santa Barbara, CA 93106}
\author{P. A. Crowell}
\affiliation{School of Physics and Astronomy, University of
Minnesota, Minneapolis, MN 55455}

\begin{abstract}
Measurements and modeling of electron spin transport and dynamics
are used to characterize hyperfine interactions in Fe/GaAs devices
with $n$-GaAs channels. Ga and As nuclei are polarized by
electrically injected electron spins, and the nuclear polarization
is detected indirectly through the depolarization of electron spins
in the hyperfine field. The dependence of the electron spin signal
on injector bias and applied field direction is modeled
by a coupled drift-diffusion equation, including effective fields
from both the electronic and nuclear polarizations. This approach is
used to determine the electron spin polarization independently of
the assumptions made in standard transport
measurements. The extreme sensitivity of the electron spin dynamics
to the nuclear spin polarization also facilitates the electrical
detection of nuclear magnetic resonance.
\end{abstract}

\maketitle
Recent observations of electrical spin injection and
detection in ferromagnet - semiconductor (FS) devices have been
confirmed by their demonstrated sensitivity to electron spin
precession. \cite{Lou1, Lou2,appel,tran,affouda,ciorga} There remain,
however, several important issues which must be resolved in order to
interpret spin transport measurements. For example, in $n$-GaAs
doped near the metal-insulator transition, spin dynamics are
influenced profoundly by hyperfine interactions. \cite{optical}
Coupling between electron and nuclear spins leads to significant
deviations from the expectations of simple drift-diffusion models.
\cite{Lou1, affouda, Salis2009} More importantly, the quantitative
determination of the electron spin polarization in a FS device using
standard transport measurements is based on assumptions about
boundary conditions and densities of states that are not
directly verifiable.  A detailed understanding of hyperfine interactions could allow
for an independent measurement of the electron spin polarization.

In this article we present measurements employing electrical
generation and detection of dynamic nuclear polarization (DNP) in
Fe$/$\emph{n}-GaAs heterostructures. Non-equilibrium spin
polarization is electrically injected into a GaAs channel, inducing
DNP.\cite{over,paget,optical} We show that the resulting dynamics
of the combined electron-nuclear spin system can be described in a
completely self-consistent manner. This approach allows for a
measurement of the electron spin polarization that is independent of
assumptions about either boundary conditions at the Fe/GaAs
interface or the electronic densities of states in the ferromagnet
or the semiconductor. Finally, we show that the spin-polarized
electrons in GaAs in combination with the ferromagnetic detector can
be used as a
sensitive probe of nuclear magnetic resonance.

\begin{figure}[t]
\centerline{\epsfbox{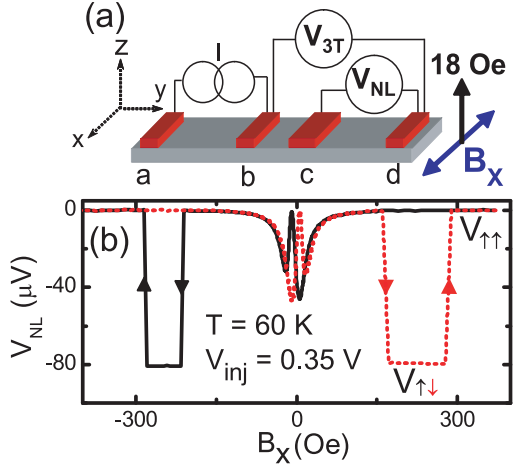}} \caption{(Color online) (a)
Schematic of the lateral spin transport device. $I$ establishes a
forward bias at contact $b$. The spin polarization in the GaAs
channel at contact $b$ or $c$ is determined by measuring the
3-terminal voltage $V_{3T}$ or non-local voltage $V_{NL}$ relative
to reference contact $d$ (drawing not to scale). (b) $V_{NL}$ for
longitudinal field $B_{x}$ swept from large positive field (black
line) and from large negative field (red dotted line).}
\label{fig:exsetup}
\end{figure}

A schematic diagram of one of our devices is shown in
Fig.~\ref{fig:exsetup}(a). \cite{Lou1,Lou2,crooker} The
heterostructures consist of a $2.5$~$\mu$m thick Si-doped n-GaAs
$(n=5\times 10^{16}$~cm$^{-3})$ channel and $5$~nm thick Fe
electrodes that are deposited epitaxially on the GaAs (100) surface.
A Schottky tunnel barrier is formed by a $n\rightarrow n^+$~$(\sim 5\times
10^{18}cm^{-3})$ GaAs transition layer \cite{hanbicki} between the
Fe and the \emph{n}-type channel. The Fe injection and detection
contacts are $7\times 50$~$\mu$m$^2$ and $4\times 50$~$\mu$m$^2$
respectively, with a center to center gap of $9.5$~$\mu$m. DNP is
observed in all heterostructures of this general design below
$80$~K. At higher temperatures, spin-polarized electrons are no
longer bound to the donor sites, thus suppressing the hyperfine
coupling.\cite{optical} The discussion in this paper is based on
data obtained at $T = 60$~K.

Measurements are performed in the non-local \cite{Lou2,
johnson} or 3-terminal \cite{Lou1,tran} configurations shown in
Fig.~\ref{fig:exsetup}(a). Under forward bias, electrons tunnel from
GaAs into the Fe injector (contact $b$). Majority spins aligned in
the $x$ direction accumulate at the injector interface \cite{Lou1}
and diffuse to the non-local detector (contact $c$). Contact $d$,
located $290~\mu$m away, serves as a voltage reference. A
longitudinal magnetic field $B_x$ is swept along the Fe easy axis
([110] direction, labeled $x$ in Fig.~\ref{fig:exsetup}(a)) at a
rate of $0.025$~Oe/s in order to ensure that the nuclear spin
polarization is in equilibrium. A small static transverse field of
$B_z = 18$~Oe is applied in the $z$ direction. This field does not
perturb the magnetization of the contacts.

\begin{figure}[t]
\centerline{\epsfbox{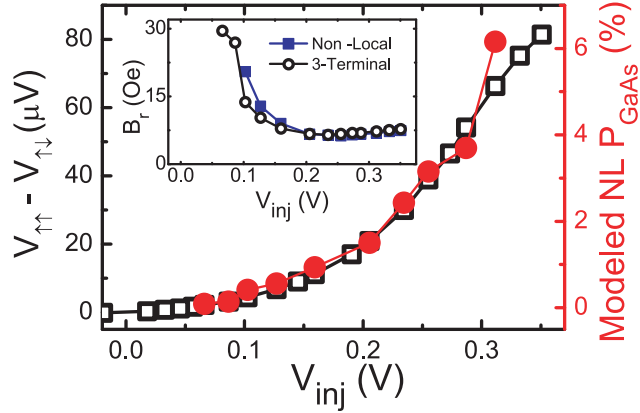}} \caption{(Color online)
$V_{\uparrow \uparrow}-V_{\uparrow\downarrow}$ (black open squares, left axis)
and modeled $P_{GaAs}$, the spin polarization averaged over the area of the
non-local detector, i.e. under
contact $c$, (red circles, right axis) as a function of injector
bias. Inset: re-polarization field $B_r$, as defined in the text,
for non-local (blue solid squares) and 3-terminal (black open circles)
measurements as a function of injector bias. } \label{fig:biasdep}
\end{figure}

We first consider
non-local spin-valve measurements. We observe jumps in the
non-local voltage $V_{NL}$ when the injector and detector
magnetizations switch between parallel and anti-parallel states,  as
shown in Fig.~\ref{fig:exsetup}(b) after subtraction of a spin-independent background.
The magnitude of the voltage
jump $V_{\uparrow \uparrow}-V_{\uparrow \downarrow}$ is proportional
to $P_{GaAs}$, where $P_{GaAs}\equiv
(n_{\uparrow}-n_{\downarrow})/(n_{\uparrow}+n_{\downarrow})$ and
$n_{\uparrow}$ and $n_{\downarrow}$ are the densities of up and down
spins respectively. Under forward bias $P_{GaAs}$ increases with
increasing $V_{inj}$, as shown in Fig.~\ref{fig:biasdep}.

\begin{figure}
\centerline{\epsfbox{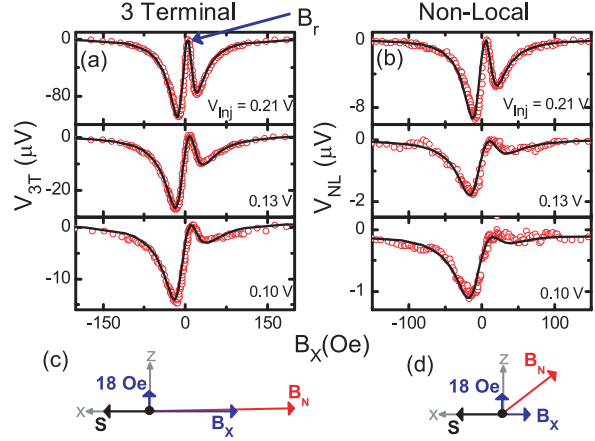}} \caption{(Color
online)3-terminal voltage $V_{3T}$ (a) and non-local voltage
$V_{NL}$ (b) as a function of applied field $B_x$ in the measurement
configuration shown in Fig.~\ref{fig:exsetup}(a) for different
injector biases $V_{inj}$. Open circles are experimental data for
$B_x$ swept from large negative field at $T=60$~K. Solid lines are
modeling results as described in the text. Schematics showing the
direction of the hyperfine nuclear field $\vec B_N$ for large $B_x$
(c), and for $B_x$ comparable to the transverse field $B_z$ (d).}
\label{fig:longsweeps}
\end{figure}

The other important aspects the data of
Fig.~\ref{fig:exsetup}(b) are the depolarization dip $(
V_{NL}\approx V_{\uparrow\downarrow})/2$ and re-polarization peak
$(V_{NL} = V_{\uparrow\uparrow})$ at low field ($|B_x| < 150$~Oe).
These features are shown on an expanded scale for several
interfacial bias voltages in Figs.~\ref{fig:longsweeps}(a) and (b).
A three-terminal measurement \cite{Lou1} is sensitive  only to
accumulation of spins parallel to the magnetization of the injector
itself and hence is a direct probe of spin precession in the
semiconductor. The fact that the field dependence of the two types
of measurements are very similar indicates that the low-field
signatures are due entirely to precession. It can be easily
verified, however, that the effective field leading to the
precessional dynamics is much larger than the applied field. This
reflects the existence of a hyperfine field due to dynamically
polarized nuclei, which has the following form:\cite{paget}
\begin{eqnarray}
\vec B_N = b_n \frac{(\vec B + b_e \vec S)\cdot \vec S (\vec B + b_e
\vec S)}{(\vec B + b_e \vec S)^2 + \xi B_{l}^{2},} \label{eq:BN}
\end{eqnarray}
where $b_n$ and $b_e$, which are both negative in GaAs,\cite{paget} 
represent effective fields due to the
polarized nuclei and electrons, $\vec S$ is the average electron
spin ($|\vec S| = 1/2$ for $P_{GaAs} = 100 \%$), $B_l$ is the local
dipolar field experienced by the nuclei, and $\xi$ parameterizes the
assisting processes which allow energy to be conserved in mutual
spin flips between electrons and nuclei. \cite{notationnote}
$|\vec B_N|$ can be as large as several Tesla in our
samples.

From Eq.~\ref{eq:BN}, we determine that at large $B_x$, $\vec B_N$
is essentially anti-aligned with $\vec S$ , as shown in
Fig.~\ref{fig:longsweeps}(c), and $\vec S$ remains polarized, with
negligible dephasing from precession. However, when $B_x$ becomes
comparable to $B_z$, $\vec B_N$ rotates towards the $z$-axis, as
shown in Fig.~\ref{fig:longsweeps}(d). $\vec S$ precesses around
$B_{N,z}$ resulting in the observed depolarization. At a smaller
field $B_r$, indicated by the arrow in Fig.~\ref{fig:longsweeps}(a),
the electron spin system becomes re-polarized, indicating that the
nuclear field has been suppressed. This phenomenon has previously
been associated with the cancellation of the magnetic field along
the spin injection axes $B_x$ by the electronic exchange field $b_e \vec S$ acting on the
nuclei.\cite{zakhar} When $B_x$ is swept
from the opposite direction the spin polarization $\vec S$ and hence
the Knight field are opposite in sign, and the sign of $B_r$
reverses, as shown in Fig.~\ref{fig:exsetup}(b).

In optical orientation experiments, the argument of the
previous paragraph has been used to determine $b_e$  by setting $B_r
= b_e S$, where $S$ is the optically generated spin accumulation.
\cite{paget, zakhar} In our experiment, we can enhance $S$ by
increasing the injector bias. As shown in
Fig.~\ref{fig:biasdep}(inset), however, $B_r$ as measured by either
the non-local or three-terminal methods clearly decreases with
increasing bias, in direct contradiction of the bias dependence of
$V_{\uparrow \uparrow}-V_{\uparrow \downarrow}$ shown in
Fig.~\ref{fig:biasdep}. This anomaly, which precludes a simple
identification of the Knight field, is due to the fact that the
electron spin polarization at small fields is reduced by precession. This fact, which was ignored in the
interpretation of the early optical orientation measurements, has a
profound influence on the electron spin dynamics at very low
fields.

To address this shortcoming as well as other aspects of our
experiments, including the spatial inhomogeneity in both the
electron and nuclear spin polarizations, we consider a more complete
model based on the drift-diffusion formalism that has been applied
widely to spin transport problems. \cite{Lou1,Lou2,johnson,crooker,jedema} The dynamics of the
injected spins in the GaAs channel are given by the following rate
equation:
\begin{eqnarray}
\frac{\partial \vec S}{\partial t} = -v_d \frac{\partial \vec
S}{\partial y}+ D \frac{\partial^2 \vec S}{\partial y^2}-\frac{\vec
S}{\tau_s}-\vec\Omega_L\times \vec S, \label{eq:dSdT}
\end{eqnarray}
where $v_d$, $D$, and $\tau_s$ are the drift velocity, diffusion
constant, and spin lifetime; $\Omega_L = g \mu_B \vec B_{tot}
t/\hbar$ is the Larmor frequency, $g=-0.44$ is the electron g-factor
in GaAs, $\mu_B$ is the Bohr magneton, and $\vec B_{tot} = \vec
B+\vec B_N$ is the sum of the external applied field and the
hyperfine field defined in Eq.~\ref{eq:BN}. In order to model the
experimental data, we solve Eq.~\ref{eq:dSdT} using the
Crank-Nicholson method \cite{num} with a one-dimensional spatial
grid. A constant spin generation rate $\dot{S}_0$ is introduced in
each cell beneath the injector. $\vec S$ and $\vec B_N$ are
calculated at each position for each time step, thus leading to a
self-consistent steady-state solution (typically after
$50$~ns).\cite{uniqueness} The non-local and 3-terminal curves as a
function of $B_x$ are determined from the spatial average of $S_x$
at the detector and injector respectively. The parameters $v_d$ and
$D$ are determined independently from Hall and resistivity
measurements. In the simulations, we set $v_d=0$ outside of the
charge current path, i.e. between the injector and detector.
$b_n,b_e, \sqrt{\xi} B_l$, and $\tau_s$ are obtained from fits of a
field sweep at one bias current and are then kept fixed while
fitting the data for other bias currents. Only $\dot{S}_0$ varies
with bias, but it is kept the same for simulations of the same bias.
As indicated by the solid curves in Figs.~\ref{fig:longsweeps}(a)
and (b), the modeling clearly reproduces the measured curves and the
re-polarization fields. \cite{fiterror}

From the modeling we determine that $b_n=-53$~kOe and $b_e=-50$~Oe,
which are, as expected, smaller than the theoretical ideal values of
$b_n = -170$~kOe and $b_e = -170$~Oe (calculated by Paget {\it et
al.}\cite{paget} for a closed electron-nuclear spin system and in
which the donor sites are always occupied by spin-polarized
electrons) but larger by a factor of 1.3 to 4 then previously
measured values for $p$-type GaAs \cite{paget} and Ge-doped GaAs.
\cite{zakhar} We find $\sqrt{\xi} B_l = 40$~Oe. $\xi$ incorporates
sample specific processes that assist nuclear spin polarization and
is given by $\xi = T_{pol}/T_1 (B/B_l)^2$.\cite{paget} Using
$T_{pol}\approx 10$~s, \cite{dyakonov} $B_l=1.45$~Oe,\cite{paget}
and a measured $T_1 = 40$~s at $B = 100$~Oe, we find $\sqrt{\xi} B_l
= 50$~Oe, which is comparable to our measured value.
\begin {figure}
\centerline{\epsfbox{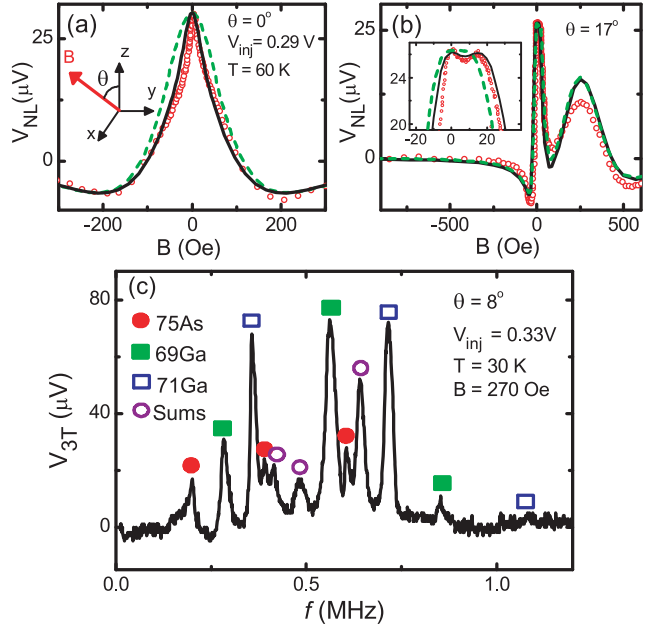}} \caption{ (Color online)
Non-local voltage field dependence: measured curves (red circles)and
modeled curves with Knight field (black line) and without Knight
field ($b_e = 0$, green dashed line) for magnetic field swept along
the z axes (Hanle) (a) and along polar angle $\theta = 17^o$ from
the z axes (oblique Hanle) (b). Modeled curves are calculated with
the same parameters used to fit data in Fig.~\ref{fig:longsweeps}.
(b) Inset: closeup of the oblique Hanle curves. (c) The
three-terminal voltage $V_{3T}$ as a function of frequency of
applied ac magnetic field. The observed resonances for each isotope
are indicated in the legend. Transitions at the sums of the
fundamental frequencies of different isotopes are also observed.}
\label{fig:hanle}
\end{figure}

The complete consistency among the different curves in
Fig.~\ref{fig:longsweeps} given a single set of parameters allows us
to use this approach to measure the bias dependence of the
spin polarization $P_{GaAs} = 2\langle|\vec{S}| \rangle$ averaged over the non-local detector.
The results extracted
from the modeling, shown in Fig.~\ref{fig:biasdep}, are in excellent
qualitative agreement with the bias dependence of $V_{\uparrow
\uparrow}-V_{\uparrow \downarrow}$. It is extremely important to
note that this measurement of $P_{GaAs}$, which is based on the
Knight field, is independent of any of the usual assumptions
underlying spin transport measurements. It is therefore of
particular interest to compare, quantitatively, with the
polarization as inferred from the non-local voltage using standard
arguments. Following Ref.\onlinecite{johnson}, $P_{GaAs}$ can be
estimated from $V_{\uparrow \uparrow}-V_{\uparrow \downarrow} = 2
\eta P_{Fe} P_{GaAs} E_f/3 e$, where $\eta$ is the interface spin
transmission efficiency, $P_{Fe}=0.4$ \cite{Soulen} is the spin
polarization of Fe, and $E_f$ is the fermi energy of GaAs, which is
assumed to be the that of a Pauli metal with an effective mass $m^* =
0.07 m_e$. From the measured $V_{\uparrow \uparrow}-V_{\uparrow
\downarrow}$ in Fig.~\ref{fig:biasdep} we have calculated $P_{GaAs}$
(for $\eta = 0.5$, which is expected based on spin-LED measurements
\cite{Adelmann} and a measured carrier density of $4.3\times 10^{16}$
~cm$^{-3}$ at $T = 60$~K) to be between 0.2$\%$ and 6$\%$ for the
injector bias range shown. This is in good agreement with the
values obtained from the analysis of the coupled electron-nuclear
spin dynamics shown in Fig.~\ref{fig:biasdep}. It is not obvious
{\it a priori} that this should be the case, since the
spin-polarized electrons responsible for the Knight shift are bound
on donors while those probed by the non-local measurement are at the Fermi level. 
Among the important
implications of this result is that the {\it magnitude} of the
density of states in the semiconductor near the Fe/GaAs interface is
not grossly different from that given by the naive Pauli model. We
have also implemented this analysis on a separate Fe/GaAs
heterostructure with the same channel doping but with $V_{\uparrow
\uparrow}-V_{\uparrow \downarrow}$ an order of magnitude lower. Fits
with the same parameters as those used to model data in
Fig.~\ref{fig:longsweeps}
yielded $P_{GaAs}$ that are in agreement with the significantly lower spin signal.

The extraordinary sensitivity of the electron spin dynamics to
hyperfine effects in the FS system was initially observed in the
distortion of the Hanle effect, in which electron spin polarization
is suppressed by precession in a transverse magnetic
field.\cite{Lou1} Fig.~\ref{fig:hanle} shows $V_{NL}$ for transverse
field swept along the $z$-axis and at a polar angle of $17^o$ from
the $z$-axis (the static field in Fig.~\ref{fig:exsetup}(a) is
eliminated). The salient features of these data are reproduced by
the model introduced above with exactly the same parameters used to
fit the data of Fig.~\ref{fig:longsweeps}. Of particular interest
are the narrowing of the Hanle curve in Fig.~\ref{fig:hanle}(a) and
the weak splitting of the zero-field peak shown in the inset of
Fig.~\ref{fig:hanle}(b). Both are due to the effect of the Knight
field. In its absence, the Hanle curves in Fig.~\ref{fig:hanle}(a)
and (b) would show the form indicated by the dashed curves. We
believe that the model's overestimation of the magnitude of the
high-field satellite peak in Fig.~\ref{fig:hanle}(b) is due to the
assumptions of purely one-dimensional diffusion as well as a uniform
current density across the injection contact.

A final indication of the strong coupling of the electron
and nuclear spin systems is provided by the detection of nuclear
magnetic resonance (NMR) when an ac magnetic field is applied by a small coil placed
over the sample.  This
is illustrated in Fig.~\ref{fig:hanle}(c), in which $V_{3T}$ is
shown as a function of frequency in a constant oblique magnetic
field at $T=30$~K. When the nuclei are off-resonance, the electron
spin polarization is suppressed by precession around $B_N$. At a resonance, the nuclei
are partially thermalized, $B_N$ is reduced, and the electron spin
polarization is restored. At low fields, this approach can be used
to observe all of the possible transitions in the $I = 3/2$ manifold
for each of the three isotopes ($^{69}$Ga, $^{71}$Ga, and $^{75}$As)
present in the sample. The observation of higher order ($\Delta m =
\pm 2, \pm 3$) transitions is likely due to dipolar interactions
(which also lead to the ``sum transitions'' from two different
isotopes) or quadrupolar coupling. \cite{strand}

In conclusion,we have demonstrated that the coupled
electron-nuclear spin dynamics in Fe/$n$-GaAs spin transport devices
can be understood quantitatively, providing an independent means of
determining the electron spin polarization. This work was supported
by NSF under DMR 0804244, the ONR MURI program, the NSF NNIN program, 
and the Japan Society for the Promotion of Science (T.K.).


\begin{thebibliography}{10}


\bibitem{Lou1}X. Lou, C. Adelmann, M. Furis, S.A. Crooker, C.J. Palmstr{\o}m, and P.A. Crowell, Phys. Rev. Lett. {\bf 96}, 176603 (2006).

\bibitem{Lou2}X. Lou, C. Adelmann, S.A. Crooker, E.S. Garlid, J. Zhang, K.S.M. Reddy, S.D. Flexner, C.J. Palmstr{\o}m, and P.A. Crowell, Nat. Phys. {\bf 3}, 197 (2007).

\bibitem{appel}I. Appelbaum, B. Huang, and D. Monsma, Nature (London) {\bf 447}, 295 (2007).

\bibitem{tran} M. Tran, H. Jaffr\`{e}s, C. Deranlot, J. M. George, A. Fert, A. Miard, and A. Lema\^{\i}tre, Phys. Rev Lett. {\bf 102}, 036601 (2009).

\bibitem{affouda}C. Awo-Affouda, O.M.J. van 't Erve, G. Kioseoglou, A.T. Hanbicki, M. Holub, C.H. Li, and B.T. Jonker, Appl. Phys. Lett. {\bf 94}, 102511 (2009)

\bibitem{ciorga}M. Ciorga, A. Einwanger, U. Wurstbauer, D. Schuh, W.
Wegscheider, and D. Weiss, Phys. Rev. B {\bf 79}, 165321 (2009).

\bibitem {Salis2009} G. Salis, A. Fuhrer, and S. F. Alvarado, arxiv:cond-mat/0908.0273

\bibitem{optical}{\it Optical Orientation}, edited by F. Meier and B.P. Zakharchenya (North-Holland, New York, 1984).

\bibitem{over}A.W. Overhauser, Phys. Rev. {\bf 92}, 411 (1953).

\bibitem{paget}D. Paget, G. Lampel, B. Sapoval, and V.I. Safarov, Phys. Rev. B {\bf 15}, 5780 (1977).

\bibitem{crooker}S.A. Crooker {\it et. al.}, {\it Science} {\bf 309}, 2191-2195 (2005).

\bibitem{hanbicki}A.T. Hanbicki {\it et. al.}, Appl. Phys. Lett. {\bf 82}, 4092-4094 (2003).

\bibitem{johnson}M. Johnson and R.H. Silsbee, Phys. Rev. Lett. {\bf 55},1790 (1985); Phys. Rev. B {\bf 37}, 5326 (1988).

\bibitem{notationnote} We adopt a notation similar to that used by Paget {\it et al.} in Ref.~\onlinecite{paget}, except that we have excluded
additional numerical prefactors for $b_n$ and $b_e$ (the leakage factor $f$ and effective donor occupancy $\Gamma$) which cannot be determined independently.

\bibitem{zakhar}B.P. Zakharchenya, V.K. Kalevich, V.D. Kul'kov, and V.G. Fle\v{i}sher, Fiz. Tverd. Tela (Leningrad) {\bf 23}, 1387-1394 (1981).

\bibitem{jedema}F. J. Jedema, H. B. Heersche, A. T. Filip, J. J. A. Baselmans, and B. J. van Wees, Nature (London) {\bf 416}, 713 (2002)

\bibitem{num}W.H. Press, S. A. Teukolsky, W.T.
Vetterling, and B.P. Flannery {\it Numerical Reciepes in C, The Art
of Scientific Computing} $2^{nd}$ ed.,(Cambridge University Press,
1992).

\bibitem{uniqueness}This approach ignores the dynamics of the nuclear spin system, for which Eq.~\ref{eq:BN} is true only in equilibrium. It will nevertheless arrive at a correct self-consistent steady-state solution for the coupled electron-nuclear spin dynamics, provided that the solution at equilibrium is unique.

\bibitem{fiterror}In fitting the data, the error in $\dot{S}_0$ is estimated to be about $20\%$. The error in nuclear field parameters is less than $20\%$.


\bibitem{dyakonov}M.I. Dyakonov and V.I. Perel, in {\it Optical
Orientation}, edited by F, Meier and B.P. Zakharchenya
(North-Holland, New York, 1984)

\bibitem{Soulen}R.J. Soulen Jr. {\it et. al.}, {\it Science} {\bf 282}, 85-88 (1998).

\bibitem{Adelmann}C. Adelmann, X. Lou, J. Strand, C.J. Palmstrom, and P.A. Crowell, Phys. Rev. B {\bf 71},121301(R) (2005).

\bibitem{strand}J. Strand, X. Lou, C. Adelmann, B.D. Schultz, A.F.
Isakovic, C.J. Palmstr{\o}m, and P.A. Crowell, Phys. Rev. B {\bf 72}
155308 (2005).


\end{thebibliography}

\end{document}